\documentclass[12pt]{article}
\usepackage{graphicx}
\usepackage{color}
\usepackage{cite}
\usepackage{url}
\usepackage{afterpage}

\newcommand{\gsim}{\lower.7ex\hbox{$\;\stackrel{\textstyle>}{\sim}\;$}}
\newcommand{\lsim}{\lower.7ex\hbox{$\;\stackrel{\textstyle<}{\sim}\;$}}

\def \beq{\begin{equation}}
\def \eeq{\end{equation}}
\def\eqref#1{(\ref{#1})}
\def\bea{\begin{eqnarray}}
\def\eea{\end{eqnarray}}
\def\jpsi{\hbox{$J\kern-0.2em/\kern-0.1em\psi$}}
\def\Y1S{\hbox{$\Upsilon(1S)$}}

\def\URLtilde{\lower0.2em\hbox{$\tilde{\phantom{a}}$}}
\def\mycomm#1{\hfill\break\strut\kern-3em{\color{red}\tt ====> #1
\color{black}}\hfill\break}

\newcount\timecount
\newcount\hours \newcount\minutes  \newcount\temp \newcount\pmhours
\hours = \time
\divide\hours by 60
\temp = \hours
\multiply\temp by 60
\minutes = \time
\advance\minutes by -\temp
\def\hour{\the\hours}
\def\minute{\ifnum\minutes<10 0\the\minutes
\else\the\minutes\fi}
\def\clock{
\ifnum\hours=0 12:\minute\ AM
\else\ifnum\hours<12 \hour:\minute\ AM
\else\ifnum\hours=12 12:\minute\ PM
\else\ifnum\hours>12
\pmhours=\hours
\advance\pmhours by -12
\the\pmhours:\minute\ PM
\fi
\fi
\fi
\fi
}

\def\monthname{\relax\ifcase\month 0/\or January\or February\or
March\or April\or May\or June\or July\or August\or September\or
October\or November\or December\else\number\month/\fi}

\def\bold#1{\setbox0=\hbox{$#1$}     \kern-.025em\copy0\kern-\wd0
\kern.05em\copy0\kern-\wd0
\kern-.025em\raise.0433em\box0 }

\begin{document}
\setcounter{footnote}{1}
\rightline{EFI 17-12}
\rightline{TAUP 3018/17}
\rightline{arXiv:1705.07691}

\begin{center}
{\Large \bf $\jpsi\,N$ photoproduction on deuterium 
\\
\vrule height 2.2ex width 0ex  
as a test for exotic baryons}
\end{center}
\bigskip

\centerline{\bf Marek Karliner$^a$\footnote{{\tt marek@proton.tau.ac.il}}
 and Jonathan L. Rosner$^b$\footnote{{\tt rosner@hep.uchicago.edu}}}
\medskip

\centerline{$^a$ {\it School of Physics and Astronomy}}
\centerline{\it Raymond and Beverly Sackler Faculty of Exact Sciences}
\centerline{\it Tel Aviv University, Tel Aviv 69978, Israel}
\medskip

\centerline{$^b$ {\it Enrico Fermi Institute and Department of Physics}}
\centerline{\it University of Chicago, 5620 S. Ellis Avenue, Chicago, IL
60637, USA}
\bigskip
\strut

\begin{center}
ABSTRACT
\end{center}
\begin{quote}
We extend a previous study of photoproduction of exotic baryon resonances
to the reaction $\gamma + d \to J/\psi + n + p$, which permits simultaneous
investigation of the reactions $\gamma + p \to P_c^+ \to J/\psi~p$ 
\hbox{($n$ spectator)} and $\gamma + n \to P_c^0 \to J/\psi~n$ 
\hbox{($p$ spectator).}  Here $P_c^+$ is an exotic baryon with quark content
$c \bar c uud$, and $P_c^0$ is its hypothetical isospin partner with quark
content $c \bar c ddu$.  We find: 
\hbox{(1) The} cross section for $J/\psi~n$ photoproduction should be equal
to that for $J/\psi~p$ photoproduction if these processes are dominated
by the photon coupling to a $c\bar c$ pair.  In that case the two processes
are equal by isospin reflection.  (2) If a $P_c^+$ candidate is a genuine
$c \bar c u u d$ resonance, its isospin partner $P_c^0 = c \bar c d d u$
should have the same mass (again by isospin reflection).  (3) In the absence
of Fermi motion, the cross section for photoproduction of $P_c$ off a deuteron
should be nearly the sum of two equal cross sections:  $\sigma(\gamma p \to
P_c^+)$ (spectator $n$) and $\sigma(\gamma n \to P_c^0)$ (spectator $p$).
(4) The effects of Fermi motion are significant. They include smearing, 
form-factor suppression and offshellness. The upshot is that the resonance is
significantly wider and the peak cross section off a deuteron is expected to be
considerably less than twice that in $\gamma p$.
\end{quote}

\smallskip

\leftline{PACS codes: 12.39.Hg, 12.39.Jh, 14.20.Pt, 14.40.Rt}
\bigskip


Two candidates for baryon resonances composed of four quarks and an
antiquark have been reported by the LHCb Collaboration \cite{LHCb},
challenging the conventional picture of baryons as exclusively three-quark
states.  The new states are a broad one with mass $4380\pm8\pm29$ MeV, width
$205\pm18\pm86$ MeV, and statistical significance $9\sigma$, and a narrower
one with mass $4449.8\pm1.7\pm2.5$ MeV, width $39\pm5\pm19$ MeV, and
statistical significance $12 \sigma$.  They are seen decaying into $\jpsi\,p$,
suggesting that their quark content is $c \bar c uud$.  The proximity of the
narrow  higher-mass state to the $\Sigma_c \bar D^*$ threshold led to its
interpretation \cite{Karliner:2015ina} as an $S$-wave $\Sigma_c \bar D^*$
molecule, while no convincing molecular interpretation was found for the
broad lower-mass state.

In order to establish either or both
$P_c^+(4380)$ and $P_c^+(4450)$ as genuine resonances
rather than kinematic enhancements, one must observe them in reactions other
than the discovery channel $\Lambda_b \to K^- \jpsi\,p$.  Several authors
\cite{Wang:2015jsa,Kubarovsky:2015aaa,Karliner:2015voa,Blin:2016dlf}
proposed the
photoproduction direct-channel reaction $\gamma\,p \to P_c^+ \to \jpsi\,p$,
for which beams at the Thomas Jefferson National Accelerator Facility
(JLAB) are uniquely suited.  In the present note we extend our proposal
to photoproduction off deuterium.  We find:

\vrule height 3.0ex width 0ex
(1) The cross section for $\jpsi\,n$ photoproduction should be equal
to that for $\jpsi\,p$ photoproduction if these processes are dominated
by the photon coupling to a $c \bar c$ pair.  In that case the two processes
are equal by isospin reflection.  (2) If a $P_c^+$ candidate is a genuine
$c \bar c u u d$ resonance, its isospin partner $P_c^0 = c \bar c d d u$
should have the same mass (again by isospin reflection).  

(3) In the absence of Fermi motion, the cross
section for photoproduction of $P_c$ off a deuteron should be nearly the
sum of two equal cross sections:  $\sigma(\gamma p \to P_c^+)$ (spectator
$n$) and $\sigma(\gamma n \to P_c^0)$ (spectator $p$).

(4) The effects of Fermi motion are significant. They include smearing,
form-factor suppression and offshellness. The upshot is that the cross
section off a deuteron is expected to be larger than, but less than
twice that in $\gamma p$.

\vrule height 3.0ex width 0ex
We now give details.
An Appendix contains material relevant to the use of a deuteron target.

It is difficult to create a heavy $c \bar c$ pair in a hadronic reaction,
whereas electromagnetic production of $c \bar c$ (as in electron-positron
annihilation) is governed only by the $c$-quark electric charge.  We are thus
justified in assuming that the reaction $\gamma + p \to X^+ \to \jpsi + p$
mainly involves the coupling of the photon to a $c \bar c$ pair, and thus the 
resonance $X^+$ has isospin $I = 1/2$ and third isospin component $I_3 = +1/2$.

Then invariance of the strong interactions under isospin reflection implies
\beq
\sigma(\gamma+p\to X^+\to \jpsi+p)=\sigma(\gamma+n\to X^0\to \jpsi+n )~,
\eeq
where $X^0$ with $I = 1/2,~I_3 = -1/2$ is the isospin partner of $X^+$.

Suppose the $\jpsi\,p$ system has a resonance $P_c^+$.  (The LHCb data
suggest two such states.)  Then since the isospin of $\jpsi$ is zero, a
corresponding resonance $P_c^0$ should show up in the $\jpsi\,n$ channel.
We can see that this general conclusion is valid also in the specialized
case of a molecule.  In that case $P_c^+(4450)$ will be composed of 
$\Sigma_c^{++} \bar D^{*-}$ with weight 2/3 and $\Sigma_c^+ \bar D^{*0}$
with weight 1/3, taking account of isospin Clebsch-Gordan coefficients.
The isospin-reflected state will be $P_c^0(4450)$, composed of $\Sigma_c^0
\bar D^{*0}$ with weight 2/3 and $\Sigma_c^+ \bar D^{*-}$ with weight 1/3.

Consider now the photoproduction of $\jpsi$ on a deuterium target.  The
shallow binding energy of the deuteron (2.2 MeV) implies that its constituents
are nearly free, with typical momenta only of order 50 MeV but with a long
tail (see Appendix).  This seemingly
small amount, however, is enough to cause considerable spread (of order 5\%)
in the incident photon energy needed to excite a narrow resonance at 4450 MeV.

To get a feeling for the size of the effect, consider first photoproduction
with real photons off a static neutron target. To excite a resonance at
4449.8 MeV, the required incident photon energy is 10,068 MeV. If instead
the momentum of the target neutron is 50 MeV parallel to the photon beam,
the required photon energy goes up to 10,618 MeV. For 50 MeV anti-parallel
momentum, it goes down to 9546 MeV. 

\begin{figure}
\begin{center}
\includegraphics[width=0.8\textwidth]{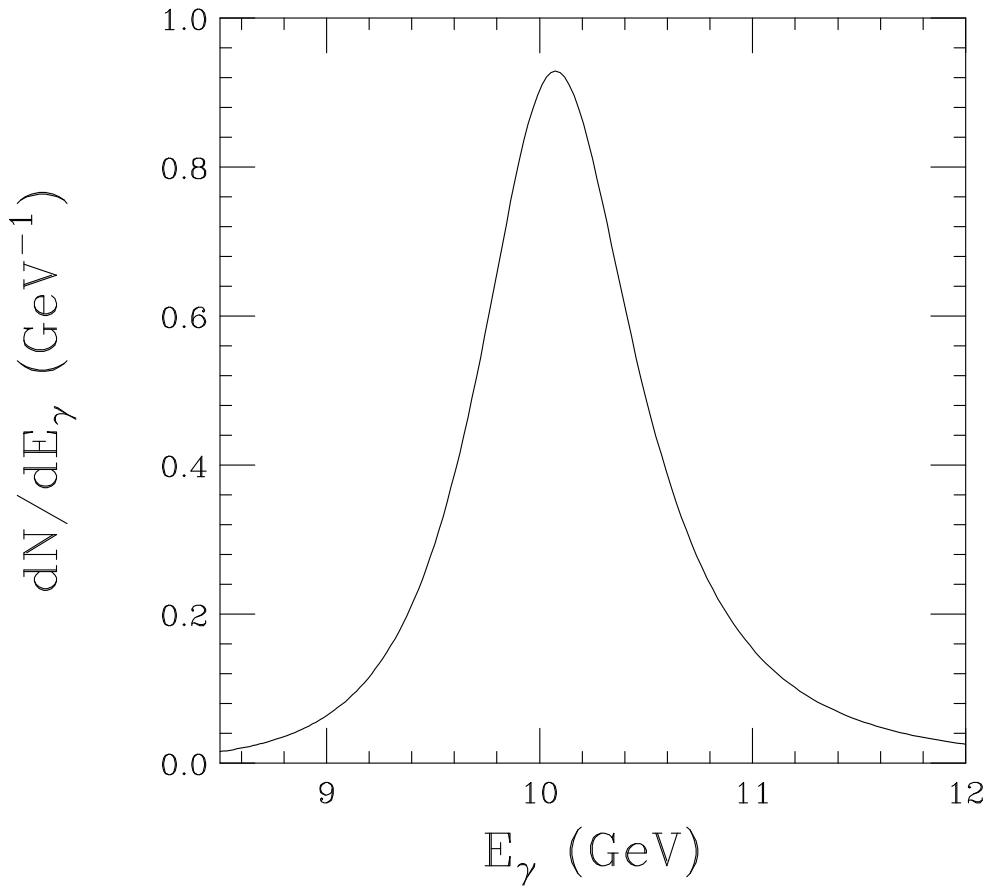}
\end{center}
\caption{Spread in incident photon energy for photoproduction of a narrow
$P_c(4450)$ on a deuteron target.  The curve is normalized so that its
integral over $E_\gamma$ is 1.
\label{fig:egam}}
\end{figure}

In Fig.\ 1 of Ref.\ \cite{Karliner:2015voa}, the effective width of a
$P_c(4450)$ photoproduced on a stationary proton target was estimated
to be 0.19 GeV.  By contrast, accounting for Fermi momentum in the deuteron
gives rise to an excitation curve with effective width of about 0.9 GeV, as
shown in Fig.\ \ref{fig:egam}.  This leads to a degradation by nearly a
factor of five of the peak resonant cross section estimated in Ref.\
\cite{Karliner:2015voa}.

In principle, detection of a recoil nucleon $N=p,n$ in the reaction $\gamma + d
\to P_c(4450) + N$ would provide a kinematic constraint reducing the needed
photon energy spread.  However, as the average recoil momentum is less than
100 MeV/$c$, one does not expect current detectors to be capable of this.

Two additional effects that might reduce the expected cross-section
as result of the Fermi motion are (a) suppression due to strong 
dependence of the form factor on $s=E_{CM}^2$ and 
(b) suppression due to the target nucleon being offshell.

(a) In exclusive processes the cross section falls off quickly
like a high power of $1/E_{CM}$, given by the ``quark counting rules"
\cite{Brodsky:1973kr}. It is therefore affected by Fermi motion. In
particular, the non-resonant cross-section will be higher if the neutron
is moving in the direction of the photon beam, because CM energy is
lower then.

(b) The kinetic energy of an onshell nucleon with momentum 100 MeV is about
5.3 MeV.  A nucleon bound inside a deuteron with nonzero Fermi momentum
must therefore be somewhat offshell, to maintain the total mass of the
deuteron unchanged. The $\gamma N$ cross-section for offshell nucleon
may be somewhat lower than for onshell nucleon.

To conclude, the photoproduction of $P_c(4450)$ on a deuteron target would
allow an important check of its nature as a genuine resonance with isospin
1/2.  Because of kinematic smearing due to the deuteron wave function, the
identification of this state is somewhat more challenging than in its
photoproduction on a hydrogen target.

We thank Or Hen, Bryan McKinnon and Stepan Stepanyan for helpful
communications.  The work of J.L.R. was supported in part by the U.S.
Department of Energy, Division of High Energy Physics, Grant No.\
DE-FG02-13ER41958.

\section*{Appendix:  Effects of deuteron wave function}

\begin{figure}
\begin{center}
\includegraphics[width=0.49\textwidth]{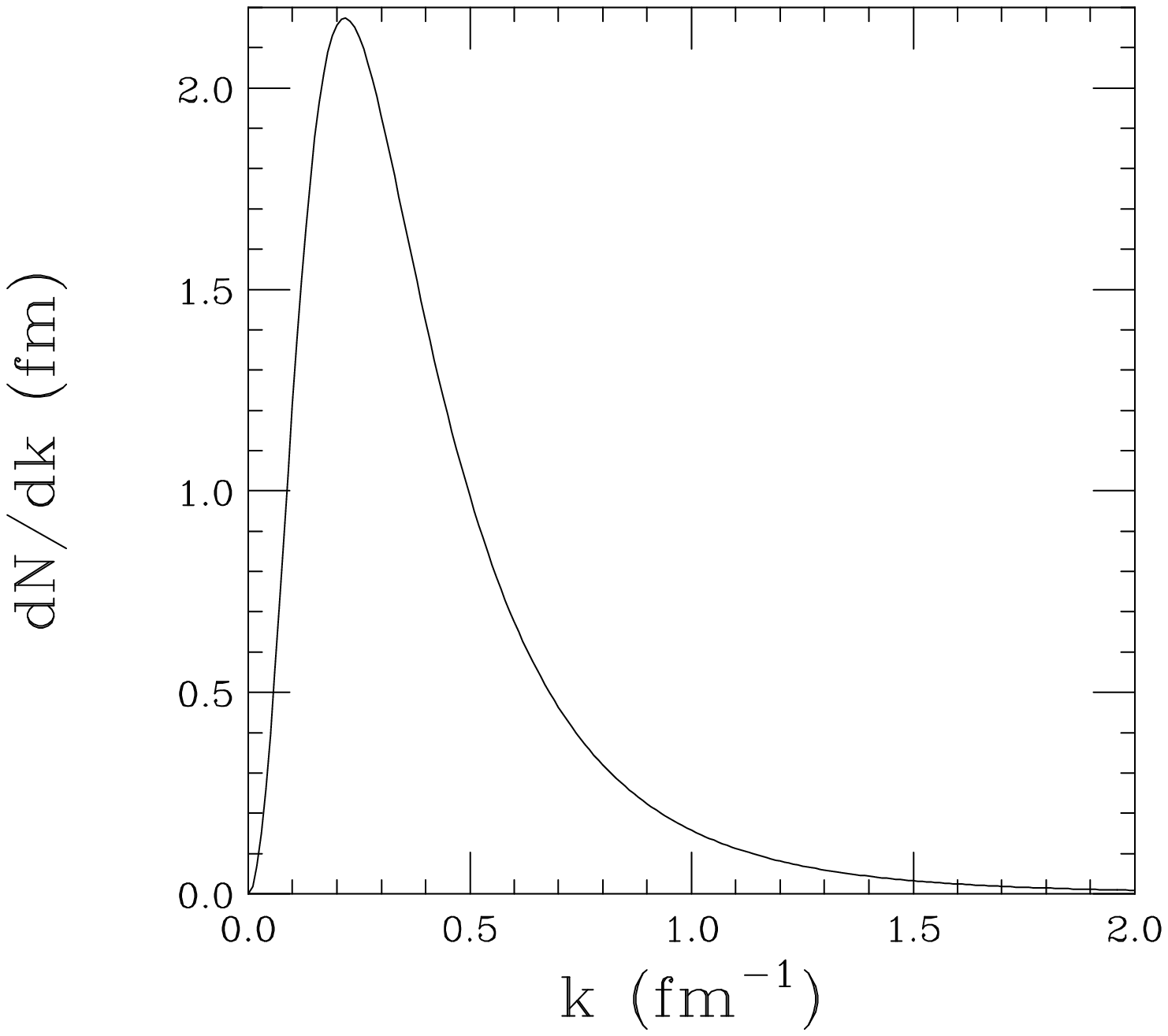}
\includegraphics[width=0.49\textwidth]{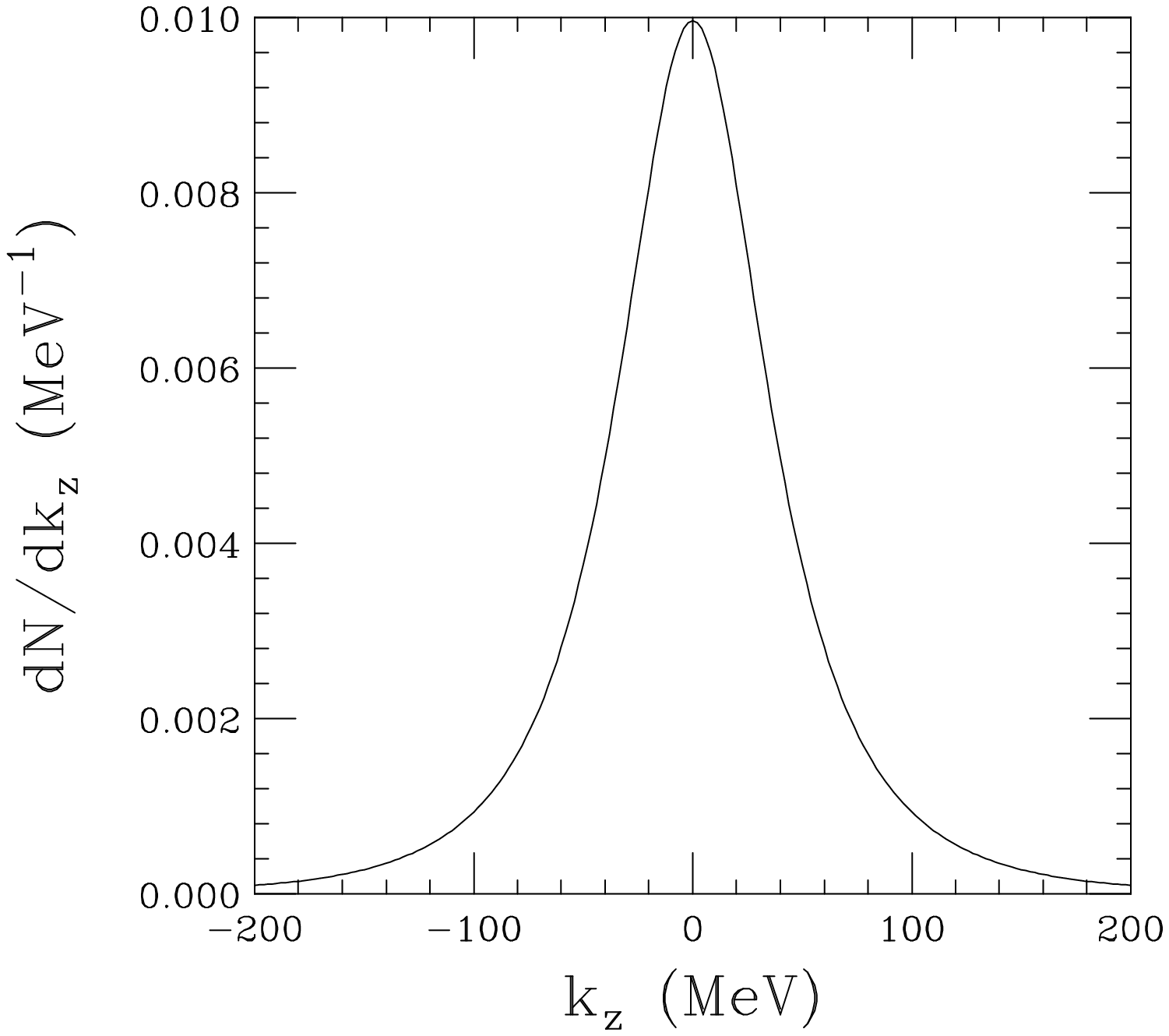}
\end{center}
\vskip-0.5cm
\caption{Left: Momentum-space distribution of nucleons inside a deuteron as
given by Eq.~\eqref{eq:psi_of_k}.  The normalization is such that the 
integral over $k$ is 1.  The peak at $k \simeq 0.22$ fm$^{-1}$ corresponds to
a momentum of 43 MeV/$c$.  Right:  Longitudinal momentum distribution of a
nucleon in a deuteron, as given by Eq.~\eqref{eq:dN_dkz}.  The normalization
is such that the integral over $k_z$ is 1.
\label{fig:wfk}}
\end{figure}

We use parameters of an S-wave Hulth\'en wave function quoted in
Ref.\ \cite{dwf}:
\beq
\psi(\vec{r}) = (c/r) [e^{-\alpha r} - e^{-\beta r}] ,
\eeq
where $\alpha = 0.232$ fm$^{-1}$, $\beta = 1.202$ fm$^{-1}$, and
$c = 0.2601$ ensures the normalization of the wave function.  This form
is easily Fourier-transformed, with the result (leaving only the dependence
on the magnitude of momentum)
\beq
\psi(k) = 4 \pi c \left[ \frac{\beta^2 - \alpha^2}{(\alpha^2+k^2)
(\beta^2+k^2)} \right]~,
\label{eq:psi_of_k}
\eeq
as shown in Fig.\ \ref{fig:wfk} (left).  The normalization of $|\psi(k)|^2$
when integrated over $4 \pi k^2 dk/(2\pi)^3$ is 1.  Equivalently, in
cylindrical coordinates,
\beq \label{eqn:cyl}
2 \pi \int k_\perp d k_\perp dk_z |\psi(k)|^2/(2\pi)^3 = 1~,
\eeq
where $k^2 = k_\perp^2 + k_z^2$.  In order to calculate the smearing in
incident photon energy needed to excite a state at fixed CM energy, we need
the distribution in {\it longitudinal} momentum $k_z$.  Performing just the
$k_\perp$ integral in (\ref{eqn:cyl}), we find
\beq
\frac{dN}{dk_z} = 2c^2 \left[ \frac{1}{\alpha^2 + k_z^2} + \frac{1}
{\beta^2 + k_z^2} + \frac{2}{(\beta^2-\alpha^2)} \log \frac{\alpha^2 + k_z^2}
{\beta^2 + k_z^2} \right]
\label{eq:dN_dkz}
\eeq
as shown in Fig.\ \ref{fig:wfk} (right).

The relation between incident photon energy and $k_z$ may be written in
simplified form as
\beq
 E_\gamma \simeq \frac{E_{CM}^2 - m_N^2}{2(m_N - k_z)}~,
\eeq
where we have neglected a small correction due to the recoil nucleon's
kinetic energy.  This allows one to transform the distribution in Fig.\
\ref{fig:wfk} (left) into that shown in Fig.\ \ref{fig:egam}.
\newpage

\begin{thebibliography}{99}


\bibitem{LHCb} R. Aaij {\it et al.} (LHCb Collaboration),
R.~Aaij {\it et al.} (LHCb Collaboration),
  Phys.\ Rev.\ Lett.\ {\bf 115} (2015) 072001 [arXiv:1507.03414 [hep-ex]].

\bibitem{Karliner:2015ina} M.~Karliner and J.~L.~Rosner,
  Phys.\ Rev.\ Lett.\ {\bf 115} (2015) 122001[arXiv:1506.06386 [hep-ph]].

\bibitem{Wang:2015jsa} Q.~Wang, X.~H.~Liu and Q.~Zhao,
  Phys.\ Rev.\ D {\bf 92} (2015) 034022 [arXiv:1508.00339 [hep-ph]].

\bibitem{Kubarovsky:2015aaa} V.~Kubarovsky and M.~B.~Voloshin,
  Phys.\ Rev.\ D {\bf 92} (2015) 031502 [arXiv:1508.00888 [hep-ph]].

\bibitem{Karliner:2015voa} M.~Karliner and J.~L.~Rosner,
  Phys.\ Lett.\ B {\bf 752} (2016) 329 [arXiv:1508.01496 [hep-ph]].

\bibitem{Blin:2016dlf} A.~N.~Hiller Blin, C.~Fernández-Ramírez, A.~Jackura,
V.~Mathieu, V.~I.~Mokeev, A.~Pilloni and A.~P.~Szczepaniak,
  Phys.\ Rev.\ D {\bf 94} (2016) 034002 [arXiv:1606.08912 [hep-ph]].

\bibitem{Brodsky:1973kr} S.~J.~Brodsky and G.~R.~Farrar,
  Phys.\ Rev.\ Lett.\ {\bf 31} (1973) 1153.

\bibitem{dwf} C. F. Perdrisat {\it et al.}, Phys.\ Rev.\ {\bf 187}
(1969) 1201.

\end{thebibliography}
\end{document}